\documentclass[amsmath,amssymb,aps,pra,superscriptaddress,twocolumn]{revtex4-1}

\usepackage{graphicx}
\usepackage{dcolumn}
\usepackage{bm}
\usepackage{siunitx}
\usepackage[normalem]{ulem}
\usepackage{placeins}

\newcommand{\ambit}{\textsc{AMB}i\textsc{T}}

\begin{document}

\title{Detection of the $5p-4f$ orbital crossing and its optical clock transition in Pr$^{9+}$ }

\author{H. Bekker}
\thanks{Present address: Department of Physics, Columbia University, 538 West 120th Street, New York, NY 10027-5255, USA}
\affiliation{Max-Planck-Institut f\"ur Kernphysik, Saupfercheckweg 1, 69117 Heidelberg, Germany}
\author{A. Borschevsky}
\affiliation{Van Swinderen Institute, University of Groningen, Nijenborgh 4, 9747 AG Groningen, The Netherlands}
\author{Z. Harman}
\affiliation{Max-Planck-Institut f\"ur Kernphysik, Saupfercheckweg 1, 69117 Heidelberg, Germany}
\author{C.~H. Keitel}
\affiliation{Max-Planck-Institut f\"ur Kernphysik, Saupfercheckweg 1, 69117 Heidelberg, Germany}
\author{T. Pfeifer}
\affiliation{Max-Planck-Institut f\"ur Kernphysik, Saupfercheckweg 1, 69117 Heidelberg, Germany}
\author{P. O. Schmidt}
\affiliation{Physikalisch-Technische Bundesanstalt, Bundesallee 100, 38116 Braunschweig, Germany}
\affiliation{Institut f\"ur Quantenoptik, Leibniz Universit\"at Hannover, Welfengarten 1, 30167 Hannover, Germany}
\author{J. R. {Crespo L\'opez-Urrutia}}
\affiliation{Max-Planck-Institut f\"ur Kernphysik, Saupfercheckweg 1, 69117 Heidelberg, Germany}
\author{J. C. Berengut}
\affiliation{Max-Planck-Institut f\"ur Kernphysik, Saupfercheckweg 1, 69117 Heidelberg, Germany}
\affiliation{School of Physics, University of New South Wales, NSW 2052, Australia}

\date{20 October 2019}

\begin{abstract}
\noindent
Recent theoretical works have proposed atomic clocks based on narrow optical transitions in highly charged ions. The most interesting candidates for searches of new physics are those which occur at rare orbital crossings where the shell structure of the periodic table is reordered. There are only three such crossings expected to be accessible in highly charged ions, and hitherto none have been observed as both experiment and theory have proven difficult. In this work we observe an orbital crossing in highly charged ions for the first time, in a system chosen to be tractable from both sides: Pr$^{9+}$. We present electron beam ion trap measurements of its spectra, including the inter-configuration lines that reveal the sought-after crossing. The proposed nHz-wide clock line, found to be at 452.334(1)~nm, proceeds through hyperfine admixture of its upper state with an E2-decaying level. With state-of-the-art calculations we show that it has a very high sensitivity to new physics and extremely low sensitivity to external perturbations, making it a unique candidate for proposed precision studies.
\end{abstract}
\maketitle

\begin{figure*}[tb]
\includegraphics[width=\textwidth]{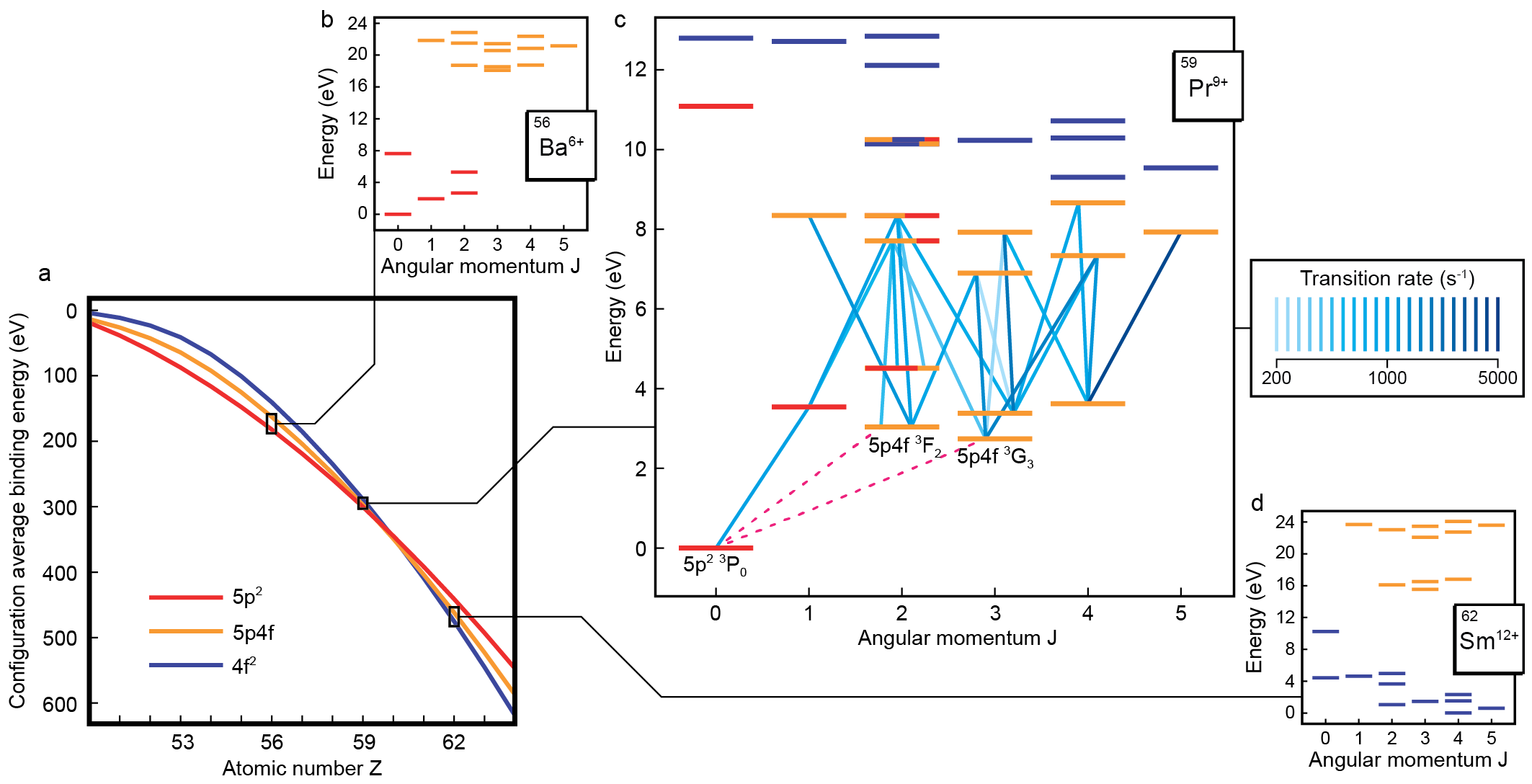}
\caption{\label{fig:levelcrossing}
a: Configuration-averaged binding energies for relevant configurations of the Sn-like (50 electron) isoelectronic sequence, as a function of atomic number $Z$. 
b -- d: Grotrian diagrams for low-lying levels of the $5p^2$, $5p4f$, and $4f^2$ configurations for $Z = 56$, 59, and 62, respectively. Pr$^{9+}$ is situated close to the configuration crossing point, and the corresponding diagram shows that inter-configuration optical transitions are allowed.
Fig 1c shows lines that were measured in the EBIT (blue). Strongly mixed $J=2$ levels are indicated with multiple colors. We identified potential clock transitions (dashed magenta lines), which are not observable in the EBIT.
}
\end{figure*}

\noindent
Current single ion clocks reach fractional frequency uncertainties $\delta\nu / \nu$ around 10$^{-18}$, and enable sensitive tests of relativity and searches for potential variations in fundamental constants~\cite{Rosenband2008, Chou2010Clock, Chou2010Relativity, Huntemann2016, Brewer2019}. These experiments could be improved by exploiting clock transitions with a much reduced sensitivity to frequency shifts caused by external perturbations. Transitions in highly charged ions (HCI) naturally meet this requirement due to their spatially compact wave functions~\cite{Schiller2007, Berengut2010}. Unfortunately, in most cases this raises the transition frequencies beyond the range of current precision lasers. While some forbidden fine-structure transitions remain in the optical range, these are generally insensitive to new physics. More interestingly, at configuration crossings due to re-orderings of electronic orbital binding energies along an isoelectronic sequence, many optical transitions between the nearly degenerate configurations can exist~\cite{Berengut2010,Berengut2012}. For the $5p$ and $4f$ orbitals, this was predicted to occur at Sn-like Pr$^{9+}$, see Fig.\ref{fig:levelcrossing}. Here, the $5p^2\ ^3$P$_0$ -- $5p4f\ ^3$G$_3$ magnetic octupole (M3) transition seems ideally suited for an ultra-precise atomic clock and searches for physics beyond the Standard Model with HCI~\cite{Berengut2012,SafronovaPRL2014,SafronovaPRA2014}, recently reviewed in~\cite{Kozlov2018}. It is highly sensitive to potential variation of the fine-structure constant, $\alpha$, and to violation of local Lorentz invariance.

\begin{figure}[!htb]
\includegraphics[width=\columnwidth]{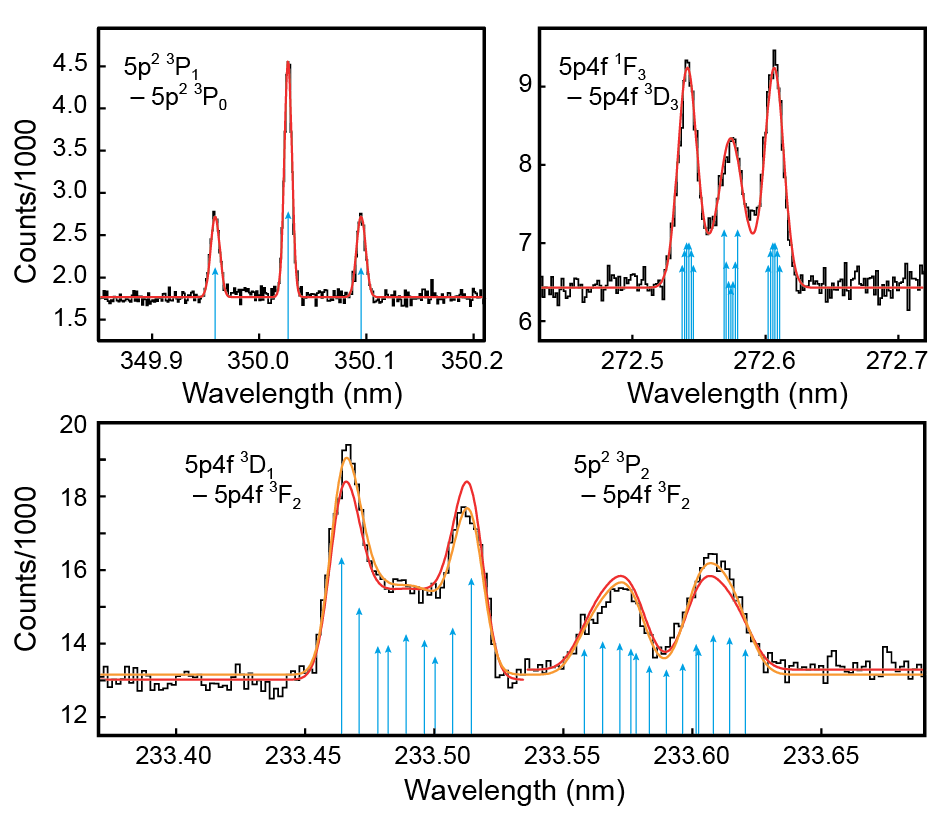}
\caption{\label{fig:spectra}Measured spectra of four distinctive Pr$^{9+}$ lines shown in black. The hyperfine Paschen-Back fit result is shown in red; blue arrows indicate the positions and relative strengths of Zeeman components (collapsing the hyperfine substructure for a clearer presentation). The orange fit in the lower panel includes the effect of Zeeman mixing of the $5p4f$ $^3$D$_1$ and $5p^2$ $^3$P$_2$ levels.}
\end{figure}

With two electrons above closed shells, Pr$^{9+}$ has a less complex electronic structure than the open $4f$-shell systems studied in previous works~\cite{Windberger2015, Murata2017, Nakajima2017}. Nonetheless, predictions do not reach the accuracy needed for finding the clock transition in a precision laser spectroscopy experiment. Instead, we measure all the optical magnetic dipole (M1) transitions with rates of at least order 100~s$^{-1}$ taking place between the fine-structure states of the $5p^2$ and $5p 4f$ configurations. Since these configurations have both even parity, strongly mixed levels exist, allowing for relatively strong M1 transitions between them. By measuring these and applying the Rydberg-Ritz combination principle, the wavelength of the extremely weak clock transition can be inferred.

\section*{Results}

The Heidelberg electron beam ion trap (HD-EBIT) was employed to produce and trap Pr$^{9+}$ ions~\cite{Crespo1999}. In this setup, a magnetically focused electron beam traverses a tenuous beam of C$_{33}$H$_{60}$O$_6$Pr molecules (CAS number 15492-48-5) which are disassociated by electron impact; further impacts sequentially raise the charge state of the ions until the electron beam energy cannot overcome the binding energy of the outermost electron. The combination of negative space-charge potential of the electron beam and voltages applied to the set of hollow electrodes (called drift tubes) trap the HCI inside the central drift tube. By suitably lowering the longitudinal trapping potential caused by the drift tubes, lower ionization states are preferentially evaporated, so that predominantly Pr$^{9+}$ ions remain trapped. Several million of these form a cylindrical cloud with a length of approximately 5~cm and radius of 200~\si{\micro\metre}. Electron-impact excitation of the HCI steadily populates states which then decay along many different fluorescent channels. Spectra in the range from 220~nm to 550~nm were recorded using a 2-m focal length Czerny Turner type spectrometer equipped with a cooled CCD camera~\cite{Bekker2018}. Exploratory searches with a broad entrance slit detected weak lines at reduced resolution. By monitoring the line intensities while scanning the electron beam energy we determined their respective charge state, finding 22 Pr$^{9+}$ lines in total; see Fig.~\ref{fig:levelcrossing}c. The charge state identification was made on the basis of a comparison between the estimated electron beam energy at maximum intensity of the lines (135(10)~eV), and the predicted ionization energy of Pr$^{9+}$ (147~eV). Here, lines from neighboring charge states appear approximately an order of magnitude weaker compared to their respective maximal intensity. Tentative line identifications were based on wavelengths and line strengths predicted from \textit{ab initio} calculations. Our Fock-space coupled cluster \cite{KalEli98} calculations were found to reproduce the spectra with average difference (theory $-$ experiment) $14\,(28)$~meV, while our \ambit~\cite{kahl19cpc} calculations (implementing particle-hole configuration interaction with many-body perturbation theory) were accurate to $-23\,(29)$~meV (see Methods for further details). Intensity ratios of the lines were compared to predictions by taking into account the wavelength-dependent efficiency of the spectrometer setup and the Pr$^{9+}$ population distribution in the EBIT. The latter was determined from collisional radiative modeling using the Flexible Atomic Code (FAC)\cite{FAC}.

For confirmation and determination of level energies at the part-per-million level, high-resolution measurements were carried out. This revealed their characteristic Zeeman splittings in the $B$~=~8.000(5)~T magnetic field at the trap center. Since the only stable Pr isotope ($A$ = 141) has a rather large nuclear magnetic moment and a nuclear spin of $I~=~5/2$, hyperfine structure (HFS) had to be taken into account to fit the line shapes. We extended the previously employed Zeeman model~\cite{Windberger2015, Bekker2018}, to include HFS in the Paschen-Back regime because $\mu_\mathrm{B}B \gg A_\mathrm{HFS}$ for the involved states. We performed a global fit of the complete data set to ensure consistent $g_J$ factors extracted from lines connecting to the same fine-structure levels. The good agreement of these with \ambit~predictions conclusively confirmed our line identifications, see Fig.~\ref{fig:spectra}, Table~\ref{tab:levels}, and Table~\ref{tab:lines}. Level energies with respect to the $5p^2\ ^3$P$_0$ ground state were determined from the wavelengths using the LOPT program~\cite{Kramida2011}, yielding those necessary to address the $5p4f\ ^3$F$_2$ and $5p4f\ ^3$G$_3$ clock states, see Table~\ref{tab:levels}. A prominent example of configuration crossing is given by the $5p4f$ $^3$D$_1$ and the $5p^2$ $^3$P$_2$ levels: their separation of 19~${\rm cm}^{-1}$ is comparable to Zeeman splitting in the strong magnetic field. This leads to a pronounced quantum interference of the magnetic substates and to an asymmetry of the emission spectra of the above fine-structure levels when decaying to the $5p4f$ $^3$F$_2$ state (see Fig. 2), also seen in e.g. the D lines of alkali metals~\cite{Tremblay1990}. The non-diagonal Zeeman matrix element between the near-degenerate fine-structure states, characterizing the magnitude of interference, was extracted by fitting the experimental line shapes, and was found to be in good agreement with \ambit\ predictions. The magnetic field strength acts in this setting as a control parameter of quantum interference. In future laser-based high-precision measurements, a much weaker magnetic field will be used. This will minimize interference and enable a better determination of field-free transition energies, and consequently of the clock lines.

\begin{table*}[!htb]
\centering
\setlength{\tabcolsep}{6pt}
\setlength\extrarowheight{4pt}
\begin{tabular}{llrrrrrrrrrr}
\hline\hline

Level & \multicolumn{7}{c}{Energy (cm$^{-1}$)} & \multicolumn{2}{c}{$g_J$} & \multicolumn{1}{c}{$A_\textrm{HFS}$} & \multicolumn{1}{c}{$q$}  \\

      & Expt. & \ambit & $\Delta$E & FSCC & $\Delta$E & 4-val~\cite{SafronovaPRA2014}  & $\Delta$E & Expt. & \ambit & \multicolumn{1}{c}{(GHz)} & \multicolumn{1}{c}{(cm$^{-1}$)} \\
\hline
$5p^2\	^3$P$_0$	&	0	&	0	&	0	&	0	&	0	&	0	&	0	&	0	&	0	&	0	&	0	\\
$5p4f\	^3$G$_3$	&	22101.36(5)	&	21368	&	-733	&	22248	&	147	&	21895(450)	&	-206	&	0.875(2)	&	0.853	&	7.771	&	69918	\\
$5p4f\	^3$F$_2$	&	24494.00(5)	&	23845	&	-649	&	24525	&	31	&	24199(370)	&	-295	&	0.889(5)	&	0.883	&	-1.688	&	64699	\\
$5p4f\	^3$D$_3$	&	27287.09(5)	&	26372	&	-915	&	27575	&	288	&	27002(570)	&	-285	&	1.136(4)	&	1.145	&	-3.857	&	74073	\\
$5p^2\	^3$P$_1$	&	28561.063(6)	&	27789	&	-772	&	28526	&	-35	&	28436(320)	&	-125	&	1.487(3)	&	1.5	&	-3.203	&	39097	\\
$5p4f\	^3$G$_4 a$	&	29230.87(6)	&	28367	&	-864	&	29482	&	251	&	29343(590)	&	112	&	1.130(3)	&	1.115	&	5.692	&	74358	\\
$5p^2\	^3$D$_2$	&	36407.48(6)	&	35550	&	-857	&	35980	&	-427	&	36217(380)	&	-190	&	1.19(1)	&	1.139	&	11.004	&	51620	\\
$5p4f\	^3$F$_3$	&	55662.43(5)	&	54852	&	-810	&	55737	&	75	&	55220(710)	&	-442	&	0.940(2)	&	0.943	&	2.568	&	110266	\\
$5p4f\	^3$F$_4$	&	59184.84(5)	&	58469	&	-716	&	59393	&	208	&		&		&	1.158(2)	&	1.161	&	2.31	&	112108	\\
$5p^2\	^3$F$_2$	&	62182.14(2)	&	61325	&	-857	&	62380	&	198	&		&		&	1.028(5)	&	1.054	&	2.224	&	101716	\\
$5p4f\	^3$G$_5$	&	63924.17(6)	&	62788	&	-1136	&	64214	&	290	&		&		&	1.202(2)	&	1.2	&	2.347	&	113269	\\
$5p4f\	^1$F$_3$	&	63963.57(6)	&	62721	&	-1243	&	64379	&	415	&		&		&	1.197(6)	&	1.226	&	0.485	&	112004	\\
$5p^2\	^3$P$_2$	&	67290.97(5)	&	66350	&	-941	&	67343	&	52	&		&		&	1.210(4)	&	1.207	&	2.331	&	98759	\\
$5p4f\	^3$D$_1$	&	67309.3(1)	&	66429	&	-880	&	67925	&	616	&		&		&	0.54(1)	&	0.5	&	-2.432	&	110679	\\
$5p4f\	^3$G$_4 b$	&	69861.70(8)	&	68528	&	-1334	&	70193	&	331	&		&		&	1.039(4)	&	1.023	&	2.62	&	111833	\\
$4f^2\	^3$F$_2$	&		&	79693	&		&	81801	&		&		&		&		&	0.813	&	1.026	&	118508	\\
$5p4f\	^1$D$_2$	&		&	80569	&		&	82657	&		&		&		&		&	0.907	&	1.555	&	123702	\\
\hline\hline
\end{tabular}
\caption{\label{tab:levels} Measured and calculated energies and Land\'e $g_J$ factors of the Pr$^{9+}$ states. Calculated magnetic-dipole hyperfine structure constants $A_\textrm{HFS}$ and sensitivities to $\alpha$-variation $q$ are also given.}
\end{table*}

\begin{figure}[tb]
\includegraphics[width=\columnwidth]{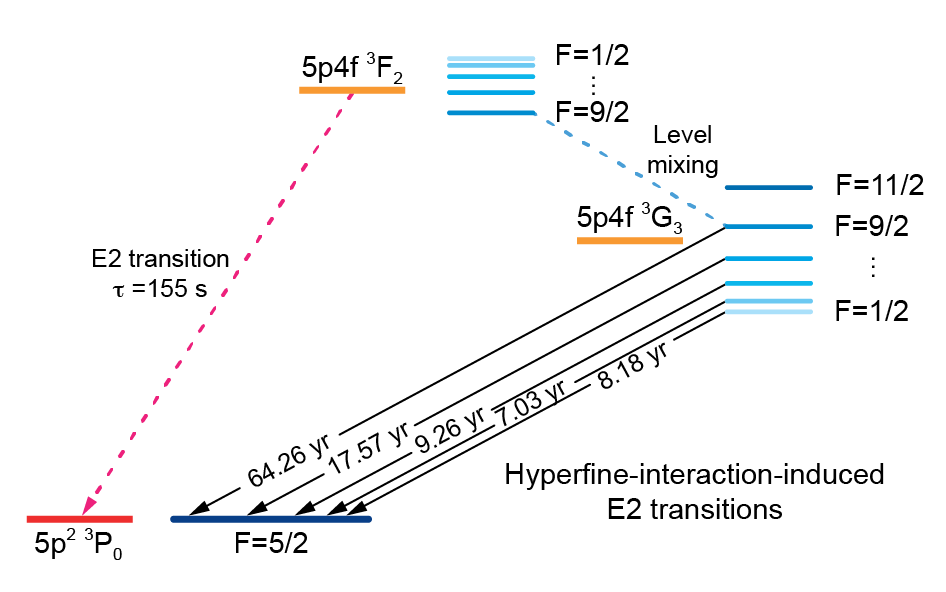}
\caption{\label{fig:HFI-E2} Schematic level diagram showing the ground and lowest two fine-structure states in Pr$^{9+}$. Levels with the same total angular momentum $F = I +J$ are admixed by the magnetic hyperfine interaction, allowing the $^3$G$_3$ clock states to decay via E2 transition rather than M3, which would take of order 10 million years to decay.
}
\end{figure}

After discovering this orbital crossing and the interesting clock transition, in the following we discuss its properties. Without HFS, the $5p4f\ ^3$G$_3$ state would decay through a hugely suppressed M3 transition with a lifetime of order 10 million years -- a 3\,fHz linewidth. However, admixture with the $5p4f\ ^3$F$_2$ state by hyperfine coupling induces much faster E2 transitions (lifetime $\sim$years) with widths on the order of~nHz, see Fig.~\ref{fig:HFI-E2}. In state-of-the-art optical clocks such transitions have been probed~\cite{Huntemann2016} and become accessible in HCI using well-established quantum logic spectroscopy techniques~\cite{schmidt_spectroscopy_2005, wolf_non-destructive_2016, chou_preparation_2017}. The $^3$G$_3\ F=11/2$ state is decoupled from the ground state but decays to the $F=9/2$ component with a lifetime of 400 years. For comparison, the $5p4f\ ^3$F$_2$ E2 transition to the ground state has a much broader linewidth of 6.4~mHz, similar to that of the Al$^+$ clock~\cite{Rosenband2007,Brewer2019}. 

Blackbody-radiation (BBR) shift is a dominant source of systematic uncertainty in some atomic clocks such as Yb$^+$\cite{Huntemann2016}. However, the static dipole polarisability $\alpha_\mathrm{S}$ (to which the BBR shift is proportional) is strongly suppressed in HCI by both the reduced size of the valence electron wavefunctions and the typically large separations between mixed levels of opposite parity. Together, these lead to a scaling $\alpha_\mathrm{S} \sim 1/Z_a^4$~\cite{Berengut2012} where $Z_a$ is the effective screened charge that the valence electron experiences. Our calculations of the static polarisability for Pr$^{9+}$ yield for the clock state of $\alpha_\mathrm{S} = 2.4$~a.u. and confirm the  expected suppression. Furthermore, the ground state polarisability is rather similar, so the differential polarisability for the clock transition is $\Delta\alpha_\mathrm{S} = 0.05$~a.u., ten times smaller than that of the excellent Al$^+$ clock transition with $\Delta\alpha_\mathrm{S} = 0.43(6)$~a.u.~\cite{Brewer2019}. An atomic clock based on Pr$^{9+}$ would therefore be extremely resilient to BBR even at room temperature.

Beyond their favourable metrological properties, HCI have been suggested as clock references for their high sensitivity to the effects of new physics~\cite{Schiller2007,Berengut2010}. Here, we investigate two particularly promising properties of the Pr$^{9+}$ clock transitions. Sensitivity to $\alpha$ variation of a transition with frequency $\omega$ is usually characterized by the parameter $q$, defined by the equation
\begin{equation}
\label{eq:q}
    \omega = \omega_0 + q x
\end{equation}
where $\omega_0$ is the frequency at the present-day value of the fine-structure constant $\alpha_0$ and $x = ({\alpha}/{\alpha_0})^2 -1$. Calculated $q$ values for Pr$^{9+}$ levels are presented in Table~\ref{tab:levels}, and compared to other proposed transitions in Table~\ref{tab:sensitivities}. The Pr$^{9+}$ M3 clock transition has a sensitivity similar to that of the 467~nm E3 clock transition in Yb$^+$, $4f^{14}6s\ ^2$S$_{1/2} \rightarrow 4f^{13}6s^2\ ^2$F$_{7/2}$ ($q=-64000$~cm$^{-1}$~\cite{dzuba08pra0}), but with opposite sign. The sign change can be understood in the single-particle model: the Yb$^+$ transition is $4f\rightarrow 6s$, while in Pr$^{9+}$ it is $5p\rightarrow 4f$, leading to opposite sign. Comparison of these two clocks would therefore lead to improved limits on $\alpha$-variation and allow control of systematics.

Invariance under local Lorentz transformations is a fundamental feature of the Standard Model and has been tested in all sectors of physics~\cite{kostelecky11rmp}. While Michelson-Morley experiments verify the isotropy of the speed of light, recent atomic experiments have placed strong limits on LLI-breaking parameters in the electron-photon sector~\cite{hohensee13prl0,pruttivarasin15nat}. The sensitivity of transitions for such studies is given by the reduced matrix element of $T^{(2)}$, defined by
\begin{equation}
\label{eq:T2}
T^{(2)}_0 = c\gamma_0 \left( \boldsymbol{\gamma}\cdot\mathbf{p} - 3\gamma_z p_z \right)
\end{equation}
where $c$ is the speed of light, ($\gamma_0$, $\boldsymbol{\gamma}$) are Dirac matrices, and $\mathbf{p}$ is the momentum of a bound electron. We find $\langle J || T^{(2)} || J \rangle = 74.2$~a.u. for the $^3$G$_3$ state, similar in magnitude to the most sensitive Dy and Yb$^+$ clock transitions, see Table~\ref{tab:sensitivities}. Again, the sign is opposite to Yb$^+$ E3, making their comparison more powerful and improving the control of potential systematic effects. Furthermore, the value compares well with other HCI~\cite{shaniv18prl}.

Future precision spectroscopy of the clock transitions will require that the internal Pr$^{9+}$ state be prepared and detected using quantum logic protocols in HCI that are sympathetically cooled in a cryogenic Paul trap~\cite{Schmoger2015, Leopold2019}. Populations calculated using FAC show that a Pr$^{9+}$ ion ejected from the EBIT ends up after a few minutes in either the $^3$P$_0$ (25\%) or $^3$G$_3$ state (75\%). We propose to employ state-dependent oscillating optical dipole forces (ODF) formed by two counter-propagating laser beams detuned by one of the trapping frequencies of the two-ion crystal with respect to each other. Electronic- and hyperfine-state selectivity may be achieved by tuning the ODF near one of the HCI resonances. If the HCI is in the target state, the ODF exerts an oscillating force onto the HCI. This displaces its motional state, which can be detected efficiently on the co-trapped Be$^+$ ion~\cite{hume_trapped-ion_2011, wolf_non-destructive_2016}, further enhanced by employing non-classical states of motion~\cite{wolf_motional_2018}.
We estimate an achievable displacement rate of tens of kHz exciting coherent states of motion for realistic parameters of laser radiation at 408~nm detuned by 1~MHz from the $^3$P$_0$ -- $5p4f\ ^3$F$_2$ transition. Similarly, motional excitation rates of a few kHz can be achieved by detuning laser radiation at 452~nm by 10~Hz from the nHz-wide $^3$P$_0$ -- ${}^3$G$_3$ clock transition. Since the ion is long-lived in both states, we distinguish between the two clock states by detuning the beams forming the ODF to additionally change the magnetic substate~\cite{chou_preparation_2017}. State selectivity can be provided by the global detuning of the ODF and the unique $g$-factor of the electronic and hyperfine states.
In case the HCI is in one of the excited hyperfine states of $^3$G$_3$, the lowest hyperfine state $F=1/2$ can be prepared deterministically by driving appropriate microwave $\pi$-pulses between $F$ states, followed by detection if the target $F$ state has been reached. The $m_F$ substate is then prepared by driving $\Delta m_F=\pm 1$ states using appropriately detuned Raman laser beams on the red sideband, followed by ground state cooling~\cite{schmidt_spectroscopy_2005, chou_preparation_2017}.

Despite the high resolution of the measured lines, the $^3$P$_0$ -- $^3$G$_3$ clock transition with expected nHz linewidth is only known to within 1.5~GHz. We can improve this by measuring the $^3$P$_0$ -- $5p4f\ ^3$F$_2$ and $^3$G$_3$ -- $5p4f\ ^3$F$_2$ transitions and performing a Ritz combination. The transitions can be found by employing again counter-propagating Raman beams that excite motion inversely proportional to the detuning of the Raman resonance to the electronic transition~\cite{wolf_non-destructive_2016}. Using realistic laser parameters similar to above, tuned near the electronic resonance, we estimate that a scan with 10~MHz steps using 10~ms probe pulses can be performed without missing the transition. Assuming that each scan point requires 100 repetitions we can scan the $\pm 2\sigma$ range in $\sim$10~minutes. By reducing the Raman laser power while extending the probe time, the resolution can be enhanced to the sub-MHz level. Once identified, quantum logic spectroscopy following~\cite{schmidt_spectroscopy_2005, rosenband_frequency_2008-1} on the clock transition can commence. Since the linewidth of the transition is significantly narrower compared to the currently best available lasers~\cite{matei_1.5_2017}, we expect laser-induced ac Stark shifts that need mitigation using hyper-Ramsey spectroscopy or a variant thereof~\cite{yudin_hyper-ramsey_2010,huntemann_generalized_2012,zanon-willette_generalized_2015,hobson_modified_2016,yudin_hyper-ramsey_2010,zanon-willette_probe_2016,yudin_generalized_2017,sanner_autobalanced_2018,beloy_hyper-ramsey_2018,yudin_combined_2018,zanon-willette_composite_2018}.

\section*{Discussion}

We have measured optical inter-configuration lines of Pr$^{9+}$, finding the $5p$ -- $4f$ orbital crossing, and thereby determined the frequency of the proposed $^3$P$_0$ -- $^3$G$_3$ clock transition with an accuracy sufficient  for quantum-logic spectroscopy at ultra-high resolution. Our state-of-the art calculations agree well with the measurements, thus we used the obtained wave functions to predict the polarizabilities of the levels and their sensitivities to new physics. These are crucial steps towards future precision laser spectroscopy of the clock transition, for which we have also proposed a detailed experimental scheme.

\section*{Methods}

\subsection{Line shape model}
In the hyperfine Paschen-Back regime, the energy shift of a fine-structure state's magnetic sublevel in an external magnetic field is given by
\begin{equation}
    E_\mathrm{PB} = g_J m_J \mu_\mathrm{B}B + A_\mathrm{HFS} m_I m_J.
\end{equation}
Hence, a transition between two fine-structure states has multiple components with energies $E_\mathrm{c}= E_0 + \Delta E_\mathrm{PB}$ with $E_0$ the transition energy without an external field, and $\Delta E_\mathrm{PB} = E'_\mathrm{PB} - E_\mathrm{PB}$. Here, and in the following, primed symbols differentiate upper states from lower states. Taking into account the Gaussian shape of individual components, the line-shape function is defined as
\begin{equation}
f(E) = \sum_{c} a_{\Delta m_J} M^2 \: \exp\left( 
-\frac{(E-E_\mathrm{c})^2}{2w^2}\right), \label{eq:fitfunction}
\end{equation}
where the sum is taken over all combinations of upper and lower magnetic sublevels. The factors $a_{\Delta m_J}$ take into account the known efficiencies of the setup for the two perpendicular linear polarizations of the light, and $w$ is the common line width which is determined by the apparatus response and Doppler broadening. The magnetic dipole matrix elements $M$ are given by
\begin{align}
    M &= \langle J, m_J, I, m_I | \boldsymbol{\mu} | J', m'_J, I', m'_I \rangle \nonumber\\
    &\propto \delta_{I, I'} \delta_{m_I, m'_I}
    \left(\begin{array}{ccc}
    J & 1 & J'\\
    -m_J & \Delta m_J & m'_J  \end{array}\right).\nonumber
\end{align}
The large parentheses denotes a Wigner $3j$-symbol. It follows that $\Delta m_J = m_J - m'_J = 0, \pm1$. In case of the asymmetric lines, the above $| J, m_J \rangle$ initial fine-structure states were transformed to the eigenstates of the Zeeman 
Hamiltonian also taking into account non-diagonal coupling.

Equation~\ref{eq:fitfunction} was fitted to each measured line with the following free parameters: $E_c$, $w$, a constant baseline value (noise floor), and overall amplitude. The results for the extracted transition energies are presented in Table~\ref{tab:lines}. The $g_J$ values of all involved states were kept as global fit parameters for the whole data set of 22 lines. Shot noise and read-out noise were taken into account for the weighting of the data points. The final uncertainties on the transition energies for each line were taken as the square root of its fit uncertainty and calibration uncertainty added in quadrature.

\begin{table}[htb]
\centering
\begin{tabular}{lllrrr}
\hline\hline
Lower & Upper & \multicolumn{3}{c}{Energy (eV)} & Rate \\
      &       & Expt. & \ambit & FSCC & (s$^{-1}$) \\
\hline
$5p^2\ ^3$D$_2$  	&	  $5p^2\ ^3$F$_2$ 	&	3.19563(1)	&	3.1957	&	3.2732	&		\\
$5p4f\ ^3$D$_3$  	&	  $5p4f\ ^3$F$_3$ 	&	3.51807(2)	&	3.5311	&	3.4916	&		\\
$5p^2\ ^3$P$_0$  	&	  $5p^2\ ^3$P$_1$	&	3.5411202(8)	&	3.4454	&	3.5368	&	1021	\\
$5p4f\ ^3$G$_4a$ 	&	  $5p4f\ ^3$F$_4$ 	&	3.713810(3)	&	3.7322	&	3.7085	&	1086	\\
$5p^2\ ^3$D$_2$  	&	  $5p^2\ ^3$P$_2$ 	&	3.829068(5)	&	3.8187	&	3.8885	&	374	\\
$5p4f\ ^3$F$_2$  	&	  $5p4f\ ^3$F$_3$ 	&	3.864391(3)	&	3.8444	&	3.8698	&	823	\\
$5p4f\ ^3$D$_3$  	&	  $5p4f\ ^3$F$_4$ 	&	3.954826(3)	&	3.9795	&	3.9449	&	675	\\
$5p4f\ ^3$G$_3$  	&	  $5p4f\ ^3$F$_3$	&	4.161043(2)	&	4.1515	&	4.1521	&	1342	\\
$5p^2\ ^3$P$_1$  	&	  $5p^2\ ^3$F$_2$	&	4.168481(3)	&	4.1579	&	4.1974	&	493	\\
$5p4f\ ^3$G$_4a$ 	&	  $5p4f\ ^3$G$_5$ 	&	4.301421(1)	&	4.2677	&	4.3062	&	4334	\\
$5p4f\ ^3$G$_4a$ 	&	  $5p4f\ ^1$F$_3$ 	&	4.306314(3)	&	4.2594	&	4.3267	&	640	\\
$5p4f\ ^3$D$_3$  	&	  $5p4f\ ^1$F$_3$ 	&	4.547299(3)	&	4.5067	&	4.5631	&	1991	\\
$5p4f\ ^3$G$_3$  	&	  $5p4f\ ^3$F$_4$ 	&	4.597753(6)	&	4.5999	&	4.6054	&	1660	\\
$5p4f\ ^3$F$_2$  	&	  $5p^2\ ^3$F$_2$	&	4.672740(8)	&	4.6469	&	4.6934	&	483	\\
$5p^2\ ^3$P$_1$  	&	  $5p^2\ ^3$P$_2$ 	&	4.80191(1)	&	4.7810	&	4.8127	&	1060	\\
$5p4f\ ^3$D$_3$  	&	  $5p^2\ ^3$P$_2$	&	4.959843(4)	&	4.9566	&	4.9306	&	762	\\
$5p4f\ ^3$G$_3$  	&	  $5p^2\ ^3$F$_2$	&	4.969391(7)	&	4.9540	&	4.9757	&	386	\\
$5p4f\ ^3$G$_4a$	&	  $5p4f\ ^3$G$_4b$ 	&	5.037586(9)	&	4.9793	&	5.0475	&	876	\\
$5p4f\ ^3$G$_3$  	&	  $5p4f\ ^1$F$_3$ 	&	5.19021(2)	&	5.1271	&	5.2236	&	252	\\
$5p4f\ ^3$D$_3$  	&	  $5p4f\ ^3$G$_4b$ 	&	5.27855(2)	&	5.2267	&	5.2840	&	878	\\
$5p4f\ ^3$F$_2$  	&	  $5p^2\ ^3$P$_2$	&	5.30616(2)	&	5.2699	&	5.3088	&	1022	\\
$5p4f\ ^3$F$_2$  	&	  $5p4f\ ^3$D$_1$	&	5.30842(2)	&	5.2797	&	5.3809	&	1199	\\
\hline\hline
\end{tabular}
\caption{\label{tab:lines} Measured lines of Pr$^{9+}$ and comparison with \emph{ab initio} theory. The rate is calculated using theoretical matrix elements from \ambit\ and experimental transition frequencies.}
\end{table}

\subsection{Fock space coupled cluster calculations}

The Fock space coupled cluster (FSCC) calculations of the transition energies were performed within the framework of the projected Dirac-Coulomb-Breit Hamiltonian \cite{Suc80}. In atomic units ($\hbar = m_e = e = 1$),
\begin{equation}
H_{\rm DCB}= \displaystyle\sum\limits_{i}h_{\rm D}(i)+\displaystyle\sum\limits_{i<j}\left(\frac{1}{r_{ij}}+B_{ij}\right).
\label{eqHdcb}
\end{equation}
Here, $h_{\rm D}$ is the one-electron Dirac Hamiltonian,
\begin{equation}
h_{\rm D}(i)=c\,\pmb{\alpha }_{i}\cdot \mathbf{p}_{i}+(\beta_{i}-1)c^2 + V_{\rm nuc}(i),
\label{eqHd}
\end{equation}
$\bm{\alpha}$ and $\beta$ are the four-dimensional Dirac matrices and $r_{ij} = |\mathbf{r}_i - \mathbf{r}_j|$.
The nuclear potential $V_{\rm nuc}(i)$  takes into account the finite size of the nucleus, modeled by a uniformly charged sphere  \cite{IshBarBin85}.  The two-electron term includes the nonrelativistic electron repulsion and the frequency-independent Breit operator,
\begin{eqnarray}
B_{ij}=-\frac{1}{2r_{ij}}\left[\pmb{\alpha }_{i}\cdot \pmb{\alpha }_{j}+(%
\pmb{\alpha }_{i}\cdot \mathbf{r}_{ij})(\pmb{\alpha }_{j}\cdot \mathbf{%
r}_{ij})/r_{ij}^{2}\right],
\label{eqBij}
\end{eqnarray}
and is correct  to second order in the fine-structure constant $\alpha$.

The calculations started from the closed-shell reference [Kr]4$d^{10} 5s^2$ configuration of Pr$^{11+}$. In the first stage the relativistic Hartree-Fock equations were solved for this closed-shell reference state, which was subsequently correlated by solving the coupled-cluster equations.  We then proceeded to add two electrons, one at at time, recorrelating at each stage, to reach the desired valence state of Pr$^{9+}$. We were primarily interested in the $5p^2$ and the $5s4f$ configurations of Pr$^{9+}$; however, to achieve optimal accuracy we used a large model space, comprised of 4 $s$, 5 $p$, 4 $d$, 5 $f$, 3 $g$, 2 $h$, and 1 $i$ orbitals. The intermediate Hamiltonian method \cite{EliVilIsh05} was employed to facilitate convergence. 

The uncontracted universal basis set \cite{MalSilIsh93} was used, composed of even-tempered Gaussian type orbitals, with exponents given by
\begin{eqnarray}
\xi _{n} &=&\gamma \delta ^{(n-1)},\text{ \ \ }\gamma =106\ 111\ 395.371\ 615 \\
\delta  &=&0.486\ 752\ 256\ 286. \nonumber
\label{eqUniversal}
\end{eqnarray}

The basis set consisted of 37 \textit{s}, 31 \textit{p}, 26 \textit{d}, 21 \textit{f}, 16 \textit{g}, 11 \textit{h}, and 6 \textit{i} functions; the convergence of the obtained transition energies with respect to the size of the basis set was verified. All the electrons were correlated.
 
The energy calculations were performed using the Tel-Aviv Relativistic Atomic Fock Space coupled cluster code (TRAFS-3C), written by E. Eliav, U. Kaldor and Y. Ishikawa. The final FSCC transition energies were also corrected for the QED contribution, calculated using the \ambit\ program.

\subsection{Polarizabilities}
The polarizabilities were also calculated using the Fock space coupled cluster method within the finite-field approach \cite{Coh65,Mon77}. We used the DIRAC17 program package \cite{DIRAC17}, as the Tel-Aviv program does not allow for addition of external fields. The v3z basis set of Dyall was used \cite{GomDyaVis10};  20 electrons were correlated and the model space consisted of 5$p$ and 4$f$ orbitals. These calculations were carried out in the framework of the Dirac-Coulomb Hamiltonian, as the Breit term is not yet implemented in the DIRAC program.

\subsection{CI+MBPT calculations}

Further calculations of energies and transition properties were obtained using the atomic code \ambit~\cite{kahl19cpc}. This code implements the particle-hole CI+MBPT formalism~\cite{berengut16pra} which builds on the combination of configuration interaction and many-body perturbation theory described in \cite{dzuba96pra} (see also~\cite{berengut06pra}). This method also seeks to solve Eqs.~(\ref{eqHdcb}) -- (\ref{eqBij}), but treats electron correlations in a very different way. Full details may be found in \cite{kahl19cpc}; below we present salient points for the case of Pr$^{9+}$.

For the current calculations, we start from relativistic Hartree-Fock using the same closed-shell reference configuration used in the FSCC calculations: [Kr]$4d^{10}5s^2$. We then create a B-spline basis set in this $V^{N-2}$ potential, including virtual orbitals up to $n = 30$ and $l = 7$.
In the particle-hole CI+MBPT formalism the orbitals are divided into filled shells belonging to a frozen core, valence shells both below and above the Fermi level, and virtual orbitals.

The CI space includes single and double excitations from the $5p^2$, $5p4f$, and $4f^2$ (``leading configurations'') up to $8spdf$, including allowing for particle-hole excitations from the $4d$ and $5s$ shells. This gives an extremely large number of configuration state functions (CSFs) for each symmetry, for example the $J = 4$ matrix has size $N=798134$. To make this problem tractable we use the emu CI method~\cite{geddes18pra0} where the interactions between highly-excited configurations with holes are ignored.

Correlations with the frozen core orbitals (including $4s$, $4p$, $3d$ shells and those below) as well as the remaining virtual orbitals ($n > 8$ or $l > 3$) are treated using second-order MBPT by modifying the one and two-body radial matrix elements~\cite{dzuba96pra}. The effective three-body operator $\Sigma^{(3)}$ is applied to each matrix element separately; to reduce computational cost it is included only when at least one of the configurations involved is a leading configuration~\cite{berengut16pra}. Finally, for the energy calculations we include an extrapolation to higher $l$ in the MBPT basis~\cite{SafronovaPRA2014} and Lamb shift (QED) corrections~\cite{flambaum05pra,ginges16jpb,ginges16pra}.

Diagonalisation of the CI matrix gives energies and many-body wavefunctions for the low-lying levels in Pr$^{9+}$. Using these wavefunctions we have calculated electromagnetic transition matrix elements (and hence transition rates), hyperfine structure, and matrix elements of $T^{(2)}$ (Eq.~\ref{eq:T2}). For all of these we have included the effects of core polarisation using the relativistic random-phase approximation (see~\cite{dzuba18pra} for relevant formulas). By contrast $q$ values (Eq.~\ref{eq:q}) were obtained in the finite-field approach by directly varying $\alpha$ in the code and repeating the energy calculation. The predicted matrix elements and sensitivities are compared to those of systems proposed or already under investigation in searches for new physics in Table~\ref{tab:sensitivities}.

\subsection{Lifetime calculations}

Direct decay of the $5p4f\ ^3$G$_3$ clock state to the ground state proceeds as an M3 transition, which is hugely suppressed and would indicate a lifetime of order 10 million years. However, the hyperfine components of the $^{141}$Pr clock state have a small admixture of $J = 2$ levels, allowing for decay via a much faster E2 transition. The rate of the hyperfine-interaction-induced decay can be expressed as a generalized E2 transition
\begin{equation}
\label{eq:R_hfs-E2}
R_\textrm{hfs-E2} = \frac{1}{15}(\omega\alpha)^5 \frac{A_\textrm{hfs-E2}^2}{2F + 1}
\end{equation}
where $F$ is the quantum number of total angular momentum of the upper state ($\mathbf{F} = \mathbf{I} + \mathbf{J}$) and the amplitude can be expressed as
\begin{multline}
\label{eq:A_hfs-E2}
A_\textrm{hfs-E2}(b \rightarrow a) = \sum_n \left[
\frac{\langle a | \hat h_\textrm{hfs} | n \rangle \langle n | Q^{(2)} | b \rangle}
{E_a - E_n} + \right.\\ \left.
\frac{\langle b | \hat h_\textrm{hfs} | n \rangle \langle n | Q^{(2)} | a \rangle}
{E_b - E_n} 
\right].
\end{multline}
Here $\hat h_\textrm{hfs}$ and $Q^{(2)}$ are the operators of the hyperfine dipole interaction and the electric quadrupole amplitude, respectively. For the clock transition the sum over intermediate states $n$ in (\ref{eq:A_hfs-E2}) is dominated by the lowest states with $J = 2$: $5p4f\ ^3$F$_2$ and $5p^2\ ^3$P$_2$. 

\subsection{MCDF calculations}

Transition energies, hyperfine structure coefficients, and $g$ factors were also evaluated in the framework of the MCDF and relativistic CI methods, as implemented in the GRASP2K atomic structure package~\cite{GRASP2K}, and were found to be in reasonable agreement with the experiment on the level of the CI+MBPT results. As a first step, an MCDF calculation was performed, with the active space of one- and two-electron exchanges ranging from the $5s$, $5p$ and $4f$ spectroscopic orbitals up to $8f$. In the second step, the active space was extended to also include the $4d$ orbitals for a better account of core polarisation effects, and CI calculations were performed with the optimized orbitals obtained in the first step. The extension of the active space in the second step has lead to 946k $jj$-coupled configurations. For a more detailed modeling of the spectral line shapes, the non-diagonal matrix elements of the hyperfine and Zeeman interactions~\cite{HFSZEEMAN} and mixing coefficients for sublevels of equal magnetic quantum numbers were also evaluated.

\begin{table}[htb]
\centering
\setlength{\tabcolsep}{6pt}
\setlength\extrarowheight{4pt}
\begin{tabular}{lclcc}
\hline\hline
Ion & Ref. & Level & $K$ & \multicolumn{1}{c}{$\langle J || T^{(2)} || J \rangle$} \\
\hline
Pr$^{9+}$ & & $5p4f\ ^3$G$_3$ & 6.32 & 74.2 \\
          & & $5p4f\ ^3$F$_2$ & 5.28 & 57.8 \\
Ca$^+$ & \cite{pruttivarasin15nat} & $3d\ ^2$D$_{3/2}$ & & 7.09\,(12) \\
       & & $3d\ ^2$D$_{5/2}$ & &9.25\,(15) \\
Yb$^+$ & \cite{dzuba16natphys,Flambaum2009} & $4f^{14}5d\ ^2$D$_{3/2}$ & 1.00 & 9.96 \\
       & & $4f^{14}5d\ ^2$D$_{5/2}$ & 1.03 & 12.08 \\
       & & $4f^{13}6s^2\ ^2$F$_{7/2}$ & -5.95 & -135.2 \\
Dy     & \cite{hohensee13prl0, Flambaum2009} & $4f^{10}5d6s\ J=10$ & 0.77 & 69.48 \\
       & & $4f^95d^26s\ J=10$ & 2.55 & 49.73 \\
Hg$^+$ & \cite{Rosenband2008, Flambaum2009}& $5d^96s^2\ ^2$D$_{5/2}$ & -2.94&  \\
\hline\hline
\end{tabular}
\caption{Overview of sensitivities to new physics of Pr$^{9+}$ clock states and those of other extensively investigated atomic systems. The relative sensitivity to variation of the fine-structure constant $K = 2q/\omega$ is given with respect to the ground state, but note that in dysprosium the fractional sensitivity of transitions between the two upper levels is many orders of magnitude higher since these levels are almost degenerate. Sensitivity to LLI is represented with the reduced matrix elements of the $T^{(2)}$ operator, given in atomic units.}
\label{tab:sensitivities}
\end{table}

\FloatBarrier

\acknowledgements
This work is part of and supported by the DFG Collaborative Research Centre \textquotedblleft SFB 1225 (ISOQUANT)\textquotedblright. JCB was supported in this work by the Alexander von Humboldt Foundation and the Australian Research Council (DP190100974). AB 
would like to thank the Center for Information Technology of the University of Groningen
for providing access to the Peregrine high performance
computing cluster and for their technical support. AB is grateful for the support of the UNSW Gordon Godfrey fellowship. POS acknowledges support from DFG, project SCHM2678/5-1, through SFB 1227 (DQ-mat), project B05, and the Cluster of Excellence EXC 2123 (QuantumFrontiers)

\section*{Author contributions}
JCB and HB conceived this work, selected the targeted ion, and wrote the manuscript. HB performed the experiment using methods devised by JRCLU, and developed the Zeeman line fitting scheme. JCB carried out \ambit\ calculations, predicted the lifetimes of the clock transitions, and identified with JRCLU the Zeeman-induced asymmetry. POS worked out the QLS scheme. AB performed FSCC and polarizability calculations; ZH carried out MCDF calculations and provided input for the asymmetric line modeling. All authors contributed to the discussions of the results and manuscript.

\section*{Data availability}
The data that support the findings of this study are available from the corresponding author upon reasonable request.

\section*{Code availability}
The \ambit~code is available at \url{https://github.com/drjuls/AMBiT}~\cite{kahl19cpc}, the LOPT program at \url{http://cpc.cs.qub.ac.uk/summaries/AEHM_v1_0.html}~\cite{Kramida2011}, the DIRAC17 package at \url{http://www.diracprogram.org}, and the GRASP2K code at \url{https://www-amdis.iaea.org/GRASP2K/}~\cite{GRASP2K}. The TRAFS-3C code is available upon reasonable request.

\bibliography{bibliography}

\end{document}